\documentclass[traditabstract]{aa}
\usepackage{txfonts}
\usepackage{graphicx}
\usepackage{natbib}
\usepackage{aalongtable}

\bibpunct{(}{)}{;}{a}{}{,}

\begin{document}

\title{One, two or three stars?  An investigation of an unusual
  eclipsing binary candidate undergoing dramatic period changes}
\author{M.~E.~Lohr\inst{\ref{inst1}}\and A.~J.~Norton\inst{\ref{inst1}}\and U.~C.~Kolb\inst{\ref{inst1}}\and D.~R.~S.~Boyd\inst{\ref{inst2}}}
\institute{Department of Physical Sciences, The Open University,
  Walton Hall, Milton Keynes MK7\,6AA, UK\\
  \email{Marcus.Lohr@open.ac.uk}\label{inst1}\and The British
  Astronomical Association, Burlington House, Piccadilly, London
  W1J\,0DU, UK\label{inst2}}
\date{Received 26 July 2013 / Accepted 30 August 2013}

\abstract {We report our investigation of
  \object{1SWASP~J234401.81$-$212229.1}, a variable with a 18\,461.6~s
  period.  After identification in a 2011 search of the SuperWASP
  archive for main-sequence eclipsing binary candidates near the
  distribution's short-period limit of $\sim$0.20~d, it was measured
  to be undergoing rapid period decrease in our earlier work, though
  later observations supported a cyclic variation in period length.
  Spectroscopic data obtained in 2012 with the Southern African Large
  Telescope did not, however, support the interpretation of the object
  as a normal eclipsing binary.  Here, we consider three possible
  explanations consistent with the data: a single-star oblique rotator
  model in which variability results from stable cool spots on
  opposite magnetic poles; a two-star model in which the secondary is
  a brown dwarf; and a three-star model involving a low-mass eclipsing
  binary in a hierarchical triple system.  We conclude that the latter
  is the most likely model.}

\keywords{stars: individual: \mbox{1SWASP~J234401.81$-$212229.1} - stars: variables: general - binaries: close - binaries: eclipsing}
\titlerunning{One, two or three stars?}
\authorrunning{M.~E.~Lohr et al.}

\maketitle

\section{Introduction}

The object \object{1SWASP~J234401.81$-$212229.1} (J2344) was
identified as a candidate W~UMa-type (contact) eclipsing binary in
\citet{norton}, primarily on the basis of light curve shape.  Using
observations from the SuperWASP archive \citep{pollacco}, a best
period of 0.21367~d was found\footnote{The corresponding object
\object{ASAS~J234402$-$2122.5} found from ASAS observations is listed
in the AAVSO International Variable Star Index as a W~UMa-type
eclipsing binary with a period of 0.2136764~d; in the ASAS Catalog of
Variable Stars as a semi-detached eclipsing binary, period 0.213678~d;
and in the Machine-learned ASAS Classification Catalog as a
$\delta$~Scuti pulsating variable with period 0.10684~d.}, giving it
immediate interest as being very close to the observed lower limit in
the period distribution of main sequence binaries of $\sim$0.2~d
\citep{ruc92}.  \citet{lohr} then found evidence of substantial period
changes in J2344, which suggested rapid period decrease on the basis
of the first four years of SuperWASP data, implying a stellar
merger within 40\,000 years at most.  However in \citet{lohr13}, which
found the object's period as 18\,461.639$\pm$0.0005~s (0.21367638~d),
more recently-available SuperWASP observations supported a subsequent
increase in period: J2344 currently appears to be undergoing dramatic
and approximately sinusoidal variations in period length.

Two other objects from the \citet{lohr13} collection of 143 SuperWASP
candidate eclipsing binaries with periods below 20\,000~s have since
been followed up spectroscopically, and in spite of relatively low S/N
spectra, were confirmed as low-mass double-lined spectroscopic and
eclipsing binaries (paper submitted to A\&A).  Therefore, with
high-resolution spectra, the prospects seemed good for confirming
J2344 as a binary system, determining its parameters and identifying
the cause of its period variation.  Observations were made with the
Southern African Large Telescope (SALT) in mid-2012, from which we
hoped to extract radial velocities for the system components.
Moreover, additional photometric observations of J2344 were made
during late 2012, with a view to measuring more recent eclipse
timings.

Here we report the surprising results of the follow-up observations:
that J2344 does not appear to be a straightforward eclipsing binary
system.  It remains a mysterious object, though certain explanations
are supported by the data while others are ruled out.  We consider
three possible models, and identify one as most plausible on current
evidence.  Whatever the true nature of this object, it is hoped that
this exploration will be of value for studies of low-mass stars and
variables in general.

\section{Observations}

\subsection{Photometry}

\begin{figure}
\resizebox{\hsize}{!}{\includegraphics{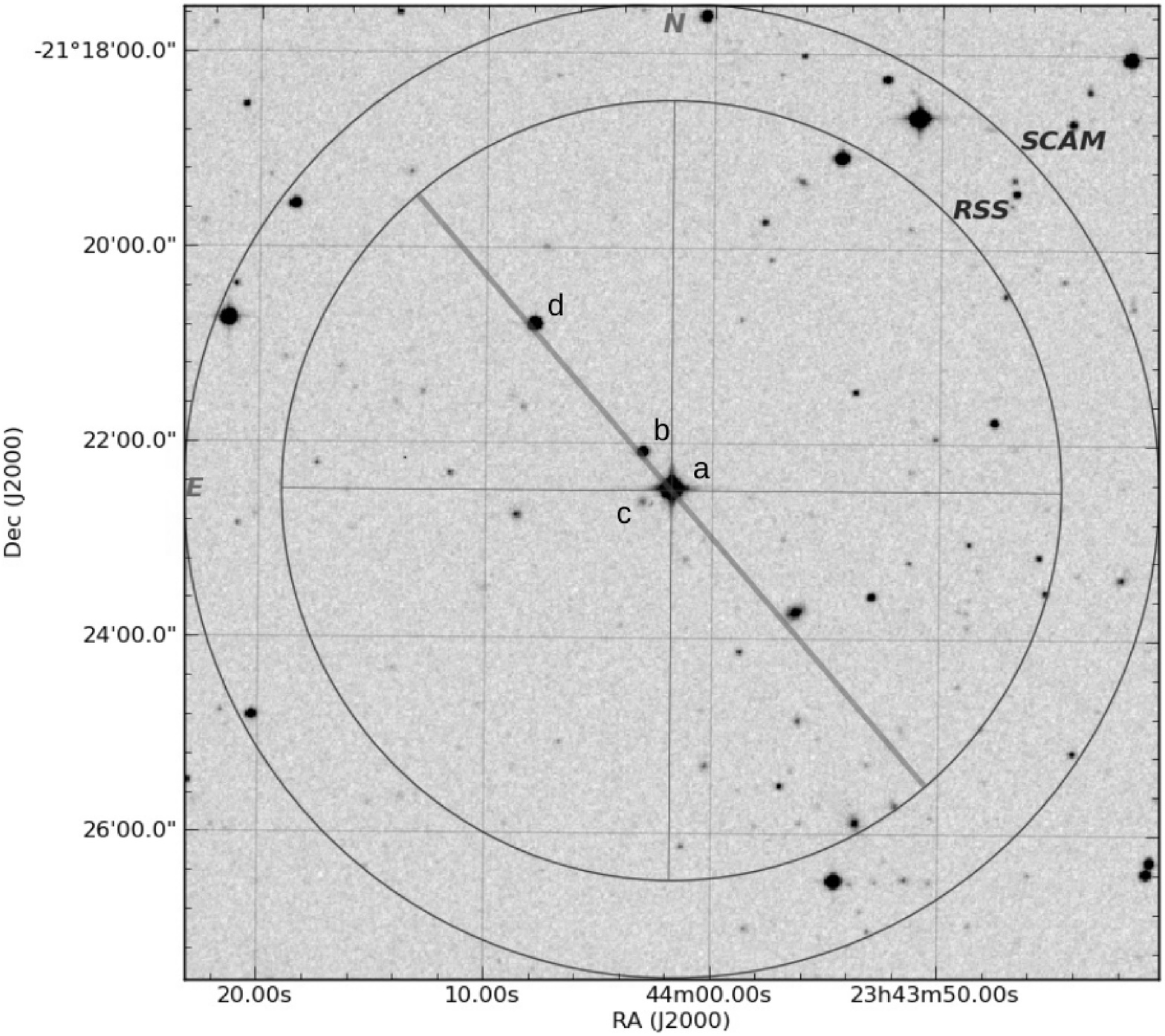}}
\caption{Local star field of J2344 (a).  The outer circle shows the
  field of view of the SALTICAM imager; the inner is that of the
  Robert Stobie Spectrograph (RSS).  The diagonal line shows the angle
  of the spectrograph slit, chosen to include potential comparison
  stars b and d.  Sources b and c would have fallen within the
  SuperWASP photometric aperture for J2344.}
\label{starfield}
\end{figure}

The SuperWASP archive contains 21727 photometric points for J2344,
obtained between 15 May 2006 and 2 August 2011.  The fluxes,
approximately corresponding to the Johnson V band, were corrected by
Sys-Rem \citep{tamuz,mazeh}, and were extracted using a
3.5~pixel-radius photometric aperture (the middle of SuperWASP's three
photometric apertures), corresponding to 47$\farcs$95.
Fig.~\ref{starfield} shows the local star field, including two nearby
sources evaluated for their possible contribution to the observed
SuperWASP light curve.  Periods and period changes were measured using
a custom IDL program, described in \citet{lohr13}, resulting in a
high-precision phase-folded light curve.

Additional photometric observations were made of J2344 and nearby
sources by D.~Boyd in the southern UK, for 0.5~h on 18 December and
1.1~h on 29 December 2012.  A 0.35~m telescope with Starlight Xpress
SXV-H9 CCD was used (pixel size 12.9~$\mu$m = 1$\farcs$2).  On 18
December, the average FWHM was 3$\farcs$6, air mass 3.4, and exposure
duration 30~s; on 29 December the corresponding values were 5$\farcs$6,
5.7 and 60~s; both nights were affected by moonlight and low-altitude
haze.  One additional eclipse timing (HJD 2456291.33132) was
determined, and the variability and magnitudes of several sources in
the vicinity of J2344 were measured, using comparison stars
GSC~06410-00829, GSC~06410-01027 and GSC~06410-00871 (magnitudes
obtained from AAVSO APASS survey).

\subsection{Spectroscopy}

53 long-slit spectra were taken for J2344 according to an automated
schedule, by duty astronomers at SALT \citep{buckley}, using the
PG1800 grating on the RSS \citep{burgh} on 1 June (16$\times$60~s), 2
July (16$\times$60~s), 1 August (4$\times$60~s) and 3 August
(16$\times$60~s; 1$\times$12~s).  By chance there was substantial
overlap in the phases covered by the second and fourth nights of
observation; a total phase coverage of $\sim$22\% was achieved.  The
slit, with a width of 0$\farcs$9, was intended to be aligned at
35.75$\degr$ to capture two nearby stars for potential comparison with
J2344 (see Fig.~\ref{starfield}), but this was achieved to varying
extents during the second, third and fourth nights, and not at all
during the first night.  A wavelength range of $\sim$5800--7100~$\AA$
was covered, to include the \ion{Na}{I}~D doublet, H$\alpha$ and a large number
of narrow metal lines characteristic of cool stars.

Primary reduction was carried out by the SALT pipeline, using the
PySALT software package\footnote{http://pysalt.salt.ac.za/}
\citep{crawford}.  This included fidelity checking, gain and
cross-talk correction, overscan bias subtraction and amplifier
mosaicking.  Master bias subtraction is not suggested for SALT data;
also, flat-fielding, cosmic-ray rejection and fringe subtraction were
not implemented in the pipeline at the time, pending calibration.
After initial attempts to use unflattened spectra, a master flat was
constructed as a median of 10 flats supplied with the August spectra,
and applied to all program images.  Spectra were then optimally
extracted using standard IRAF routines (which effectively cleaned out
cosmic rays), and calibrated using neon arc lamp exposures.  A
resolution of $\sim$0.41~$\AA$ per pixel was obtained.

In the apparent absence of visible line splitting or shifting in the
spectra, no data-internal determination of phase was possible.
Therefore phases were assigned to the spectra using a SuperWASP
ephemeris in combination with D.~Boyd's more recent eclipse timing.
The source's spectral type was confirmed by cross-correlation using
the IRAF task FXCOR, with comparison templates drawn from the
Indo-U.S. Library of Coud\'{e} Feed Stellar Spectra \citep{valdes},
which uses a comparable resolution (0.44~\AA) and wavelength range
(3460--9464~\AA).  Cross-correlation with a program spectrum of phase
0 was used to measure radial velocities (RVs), since the assumed two
component spectra would be coincident during the primary eclipse.

\section{Results}

\begin{figure}
\resizebox{\hsize}{!}{\includegraphics{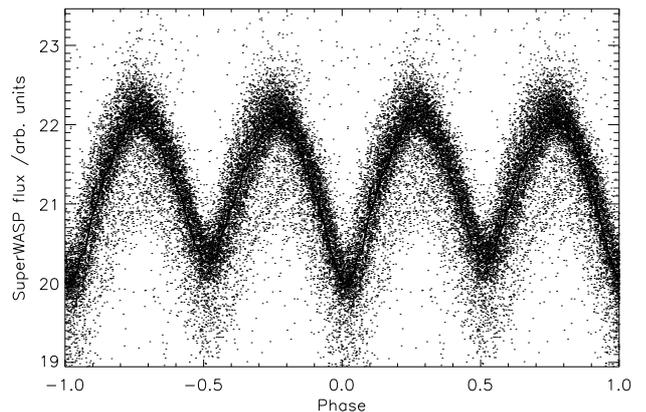}}
\caption{SuperWASP light curve for J2344, folded at period of
  18\,461.639~s, with binned mean curve overplotted.}
\label{lc}
\end{figure}

Fig.~\ref{lc} shows J2344's light curve, folded at its optimal period
of 18\,461.639~s.  There is a small but consistent difference in the
depths of primary and secondary minima, and continuous light
variation, explaining its preliminary identification as an eclipsing
binary in thermal contact.  However, we may note the small amplitude
of variation relative to the maximum or `out-of-eclipse' flux level of
$\sim$22 units ($\sim$11.6~V mag): only about 1/11 of the light is
lost during the assumed primary eclipse.  This would imply a low angle
of inclination of the system, a low mass ratio, and/or a third light
entering the aperture.

\begin{figure}
\resizebox{\hsize}{!}{\includegraphics{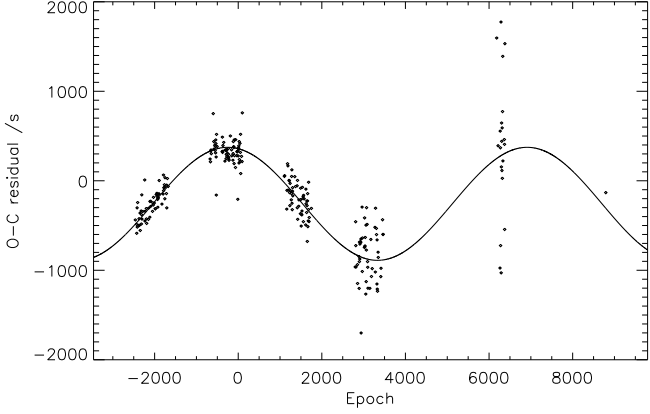}}
\caption{O$-$C diagram for J2344, following subtraction of best linear
  fit.  Best sinusoidal fit is overplotted (meta-period 4.19~y;
  $\chi^2=2.82$).  Uncertainties on individual points are not shown
  for clarity, but are typically of the order of $\pm$100~s.  The
  final point, around epoch 9000, corresponds to the independent observation of
  D.~Boyd, and has an uncertainty of $\pm$60~s.}
\label{oc}
\end{figure}

\citet{lohr13} discusses the period changes we observed in J2344, and
its Figs.~5 and 6 illustrate the best linear and quadratic fits to the
data.  The first four years suggested a rapid period decrease
(reflected in an O$-$C parabola opening downwards), but the most
recent year of SuperWASP data conflicted with this model
($\chi^2=17.83$); assuming a half-cycle error in the primary minimum
fits of the final year's data improved the quadratic fit
($\chi^2=10.86$) but left Year 4 now visibly discrepant.  Here,
Fig.~\ref{oc} gives the best sinusoidal fit to the SuperWASP data and
the additional observation of a primary minimum from 2012; this
provides an optimal model ($\chi^2=2.82$), and suggests a meta-period
for J2344 of 4.19$\pm$0.04 years, with an amplitude of 631$\pm$11~s.
The observed times of minima are listed in Table~\ref{mintab}.

\begin{figure}
\resizebox{\hsize}{!}{\includegraphics{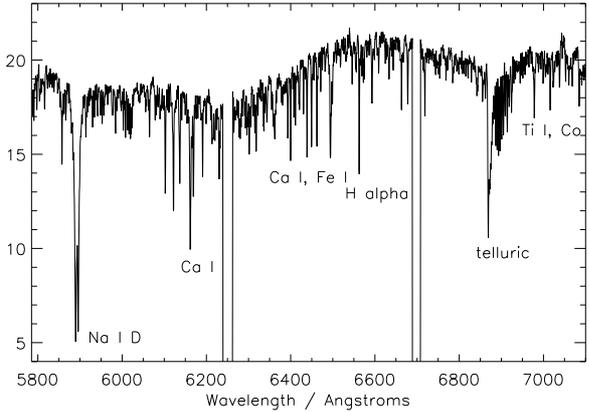}}
\caption{SALT spectrum at phase 0.588.  The gaps around 6250 and 6700~$\AA$ correspond to the boundaries of the three CCDs.}
\label{spec}
\end{figure}

\begin{figure}
\resizebox{\hsize}{!}{\includegraphics{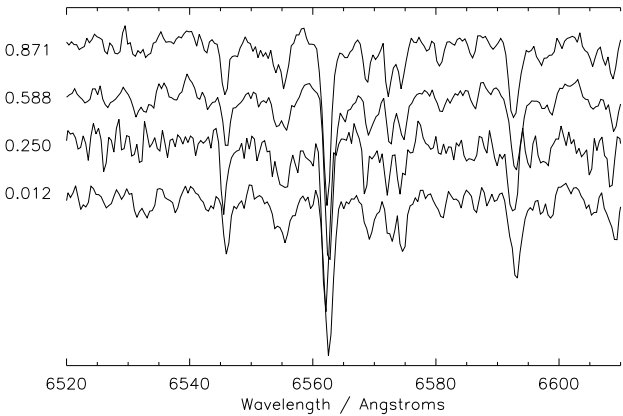}}
\caption{Sections of SALT spectra around H$\alpha$ line at
  6562.8~$\AA$, selected from the four nights of observation, and
  covering as wide a phase range as possible (phases shown on left).}
\label{specarray}
\end{figure}

Fig.~\ref{spec} shows an example full extracted and calibrated
spectrum for J2344.  The best-matching comparison spectra were around
K5V (temperatures between 4000 and 4500~K), achieving
cross-correlation peak heights in excess of 0.95.  However, to our
surprise, little to no splitting or even shifting of the many
well-defined absorption lines observed was apparent to the eye, as
would be expected in a close, short-period eclipsing binary
(Fig.~\ref{specarray}).  Only for spectra near phase 0.25 is there any
suggestion of a leftward shift, and unfortunately, these spectra are
by far the faintest of all four sets of observations, reducing their
reliability.  Moreover, an approximate light curve extracted from the
spectra themselves (by fitting a spline to each continuum and evaluating
it at a given wavelength) did not reflect the SuperWASP light curve at
all, being apparently dominated by systematic effects.

\begin{table*}[!]
\caption{Summary of spectroscopic observations and RVs
  for J2344}
\label{rvtable}
\centering
\begin{tabular}{l l l l | l l l l}
\hline\hline
HJD & Phase & RV & $\delta$ RV & HJD & Phase & RV & $\delta$ RV\\
$-2450000$ & & (km~s\textsuperscript{-1}) & (km~s\textsuperscript{-1})
& $-2450000$ & & (km~s\textsuperscript{-1}) & (km~s\textsuperscript{-1})\\
\hline
6079.6205 & 0.197 & 0.1 & 2.2 & 6110.5461 & 0.928 & -0.2 & 1.7\\
6079.6214 & 0.201 & -1.1 & 1.5 & 6110.5480 & 0.937 & -0.6 & 1.3\\
6079.6222 & 0.205 & -2.0 & 1.8 & 6110.5489 & 0.941 & -0.8 & 1.5\\
6079.6231 & 0.209 & -0.6 & 1.9 & 6110.5498 & 0.945 & -2.0 & 1.3\\
6079.6252 & 0.219 & 1.0 & 1.8 & 6110.5506 & 0.949 & -1.3 & 1.2\\
6079.6261 & 0.223 & 1.3 & 2.0 & & & & \\ 
6079.6270 & 0.228 & 2.9 & 2.6 & 6141.4557 & 0.584 & 3.7 & 1.2\\ 
6079.6278 & 0.231 & 1.0 & 2.0 & 6141.4566 & 0.588 & 3.9 & 1.5\\ 
6079.6300 & 0.242 & 2.7 & 2.4 & 6141.4575 & 0.593 & 2.6 & 1.3\\ 
6079.6309 & 0.246 & 1.4 & 1.8 & 6141.4583 & 0.596 & 3.2 & 1.3\\
6079.6317 & 0.250 & -0.2 & 2.3 & & & & \\
6079.6326 & 0.254 & 1.3 & 2.1 & 6143.4463 & 0.900 & 4.8 & 1.3\\
6079.6348 & 0.264 & 0.9 & 2.0 & 6143.4471 & 0.904 & 3.9 & 1.4\\
6079.6357 & 0.268 & 0.2 & 1.9 & 6143.4480 & 0.908 & 4.9 & 1.2\\
6079.6366 & 0.272 & 3.2 & 1.9 & 6143.4489 & 0.912 & 5.5 & 1.2\\
6079.6375 & 0.277 & 4.3 & 1.8 & 6143.4534 & 0.933 & 5.7 & 1.0\\
 & & & & 6143.4543 & 0.938 & 5.3 & 1.4\\
6110.5340 & 0.871 & 0.3 & 1.7 & 6143.4552 & 0.942 & 5.2 & 1.2\\
6110.5349 & 0.876 & 0.8 & 2.1 & 6143.4560 & 0.946 & 5.3 & 0.9\\
6110.5358 & 0.880 & 1.0 & 1.5 & 6143.4607 & 0.968 & 5.0 & 1.0\\
6110.5367 & 0.884 & 1.1 & 2.0 & 6143.4616 & 0.972 & 2.9 & 0.9\\
6110.5387 & 0.893 & 1.5 & 1.8 & 6143.4625 & 0.976 & 3.4 & 1.1\\
6110.5396 & 0.898 & 1.3 & 1.4 & 6143.4640 & 0.983 & 4.7 & 1.3\\
6110.5405 & 0.902 & 2.0 & 1.4 & 6143.4647 & 0.986 & 3.2 & 0.9\\
6110.5413 & 0.906 & 1.3 & 1.6 & 6143.4675 & 0.999 & 0.0 & 0.0\\
6110.5434 & 0.915 & 1.2 & 1.5 & 6143.4684 & 0.004 & -1.1 & 1.1\\
6110.5443 & 0.920 & 0.6 & 1.6 & 6143.4693 & 0.008 & -2.8 & 0.9\\
6110.5452 & 0.924 & 0.7 & 1.6 & 6143.4702 & 0.012 & -4.6 & 0.9\\
\hline
\end{tabular}
\end{table*}

\begin{figure}
\resizebox{\hsize}{!}{\includegraphics{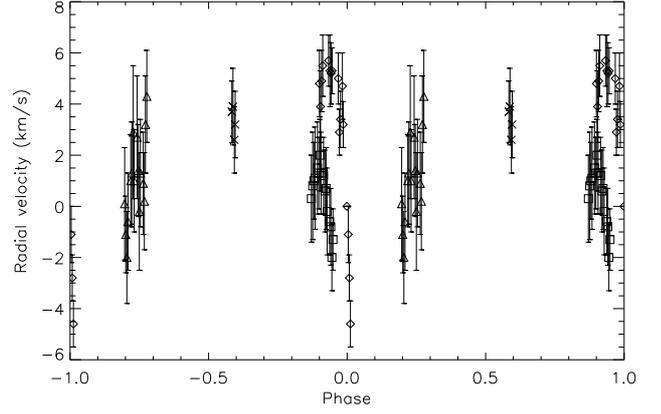}}
\caption{RV curve for J2344.  First night's observations
  are plotted with triangles; second night: squares; third night:
  crosses; fourth night: diamonds.}
\label{rv}
\end{figure}

Table~\ref{rvtable} gives the heliocentric times, estimated phases and
RVs for J2344's spectroscopic observations.  Only one
clear cross-correlation peak was seen for each spectrum, rather than
the two that we would expect for an eclipsing binary.  (Repetition of
the measurements using a template K5V spectrum produced very similar
results, apart from a systematic offset due to relative centre-of-mass
system velocities.)  Fig.~\ref{rv} plots the resulting RV
curve against phase.  We may note that the amplitude of variation is
very small: only $\pm$5~km~s\textsuperscript{-1}, where we might
expect values of tens or hundreds of km~s\textsuperscript{-1}.  Also,
such trends as are suggested over the orbital cycle do not seem to
correspond to expected velocity changes for either component of a
spectroscopic binary: some sort of maximum appears around phase 0.9
during the region of overlapping phase coverage, but this is when we
expect the primary's RV to be decreasing, and the secondary's to be
increasing; phase 0.25 should correspond to the primary's minimum RV
and the secondary's maximum, but actually shows a rising trend in our
plot.

\section{Discussion}

Our spectroscopic results were not as expected given the photometric
data for J2344.  Its SuperWASP light curve (and indeed, the ASAS light
curve of the corresponding source) strongly suggests a very short
period eclipsing binary in contact configuration, like those of
1SWASP~J150822.80$-$054236.9 and 1SWASP~J160156.04+202821.6, which
were recently confirmed as spectroscopic double-lined and eclipsing
binaries, on the basis of fairly low-resolution INT spectra, and
modelled as contact systems (paper submitted to A\&A).  However, the
SALT spectra for J2344 showed little to no evidence for line splitting
or shifting, being instead strongly consistent with a single, stable
mid-K star.

\begin{figure}
\resizebox{\hsize}{!}{\includegraphics{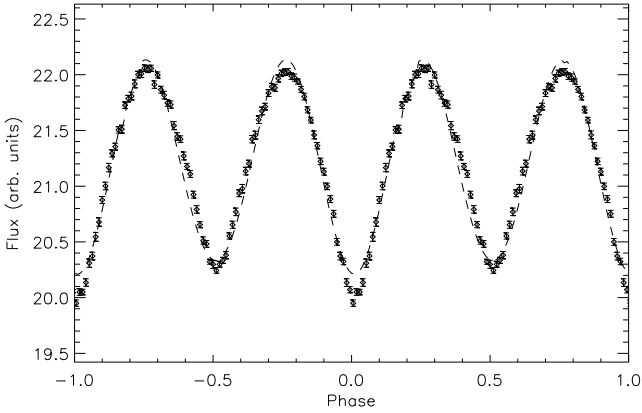}}
\caption{Best light curve fit for PHOEBE model 1 of eclipsing binary
assuming primary has mass consistent with K5V star, and secondary is
massive enough to burn hydrogen ($i=44\degr$, $q=0.11$,
$M_1=0.73M_{\sun}$, $M_2=0.08M_{\sun}$).  Binned input SuperWASP light
curve shown with diamonds and uncertainty bars; fit with dotted line.}
\label{lcfit1}
\end{figure}

\begin{figure}
\resizebox{\hsize}{!}{\includegraphics{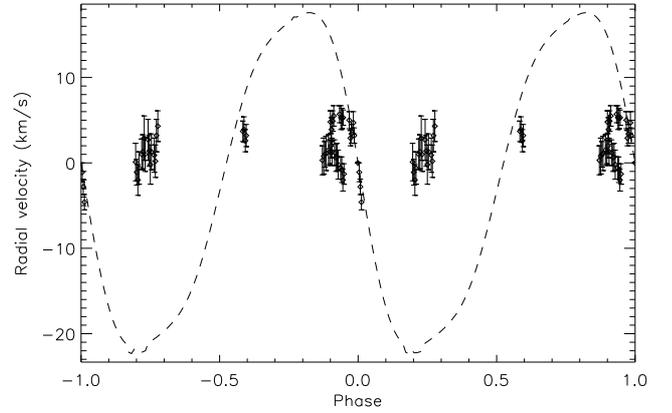}}
\caption{Best primary RV curve fit for PHOEBE model 1 (parameters
as for Fig.~\ref{lcfit1}).  SALT RV curve shown with
diamonds and uncertainty bars; fit with dotted line.}
\label{rvfit1}
\end{figure}

\begin{figure}
\resizebox{\hsize}{!}{\includegraphics{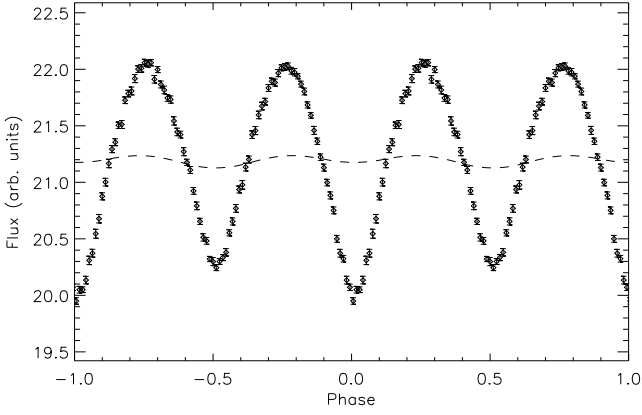}}
\caption{Best light curve fit for PHOEBE model 2 of eclipsing binary
assuming primary has mass consistent with K5V star, and secondary is
massive enough to burn hydrogen ($i=10\degr$, $q=0.11$,
$M_1=0.73M_{\sun}$, $M_2=0.08M_{\sun}$).}
\label{lcfit2}
\end{figure}

\begin{figure}
\resizebox{\hsize}{!}{\includegraphics{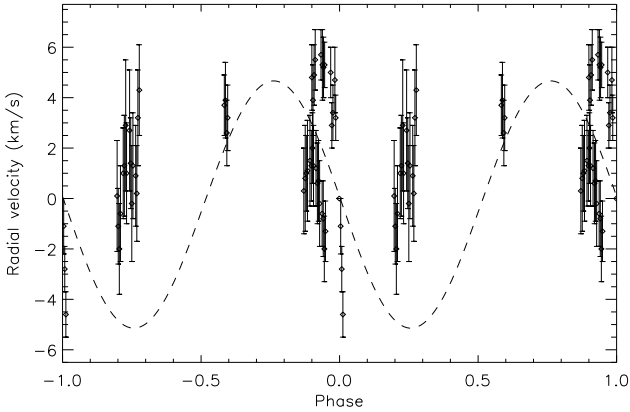}}
\caption{Best primary RV curve fit for PHOEBE model 2 (parameters as for
Fig.~\ref{lcfit2}).}
\label{rvfit2}
\end{figure}

To confirm our impression of the inconsistency of the photometric and
spectroscopic results, modelling was carried out using the eclipsing
binary modelling software PHOEBE \citep{prsa}, built upon the code of
\citet{wildev}.  Figs.~\ref{lcfit1} to \ref{rvfit2} illustrate the
best light curve and RV curve fits for two models: one optimising
light curve fit and the second optimising RV curve fit.  Input
parameters of semi-major axis and mass ratio were constrained by the
requirements that the more massive star in the assumed binary be
consistent with a K5V spectrum, and its companion be massive enough to
burn hydrogen, so that the system contains two main sequence stars.
It may be seen that, with the minimum mass ratio $q=0.11$, the
observed light curve can be tolerably reproduced with a moderate angle
of inclination $i=44\degr$ (Fig.~\ref{lcfit1}), but the corresponding
primary RVs are then about three times larger than observed
(Fig.~\ref{rvfit1}).  However, if $i$ is reduced far enough to bring
the modelled RV curve into the observed range (Fig.~\ref{rvfit2}), the
corresponding light curve model has far too small an amplitude
(Fig.~\ref{lcfit2}).  Higher mass ratios fail to reproduce either
light curve or RV curve, at any angle of inclination.  Therefore we
conclude that the observed photometry and spectroscopy, taken
together, are incompatible with any low-mass eclipsing binary composed
of main sequence stars.

One initial explanation considered was that J2344 was not in fact the
source of the light variability observed by SuperWASP (and ASAS).
Objects b and c, shown in Fig.~\ref{starfield}, are close enough to
J2344 to have fallen within the same SuperWASP photometric aperture.
Could one of these be the expected eclipsing binary?  Object b was
captured within the SALT slit during three nights of observations, so
its spectrum was also extracted and reduced.  Although much fainter
and noisier than J2344's spectrum, the strongest absorption lines
(H$\alpha$ and \ion{Na}{I}~D) were consistently visible, but showed no
signs of shifting or splitting.  Also, the SuperWASP archive contains
a light curve for object b (1SWASP~J234403.11-212205.8) which we
obtained and analysed; in the 3.5 pixel aperture it showed similar
variability to J2344, but in the smaller 2.5 pixel aperture, which
should have excluded most of its neighbour's flux, its variability was
less periodic, while J2344's light curve exhibited the same periodic
behaviour even in the small aperture.  Finally, D.~Boyd confirmed from
his 18 December 2012 observations of the field of view that object b
had an (unfiltered) magnitude of 15.46$\pm0.21$, corresponding to a
SuperWASP flux variation of only about $\pm0.3$~units: far smaller
than J2344's amplitude of $\pm1$~unit.  He also noted that object
c did not surpass his sky background level of 16.5--17.0~mag; it
therefore would have been too faint to be detectable by SuperWASP,
with its range of $\sim$8--15~$V$~mag.  We must conclude, then, that
J2344 really is the source of the periodic photometric variation
observed.

What, then, might explain an object with the light curve of a
short-period binary but the spectrum of a single star?  Moreover, what
is the source of its dramatic period changes, with their 4.19~y
meta-cycle?  We have explored three physical models which provide
potential explanations for these observations.

\subsection{One-star model}

\begin{figure}
\resizebox{\hsize}{!}{\includegraphics{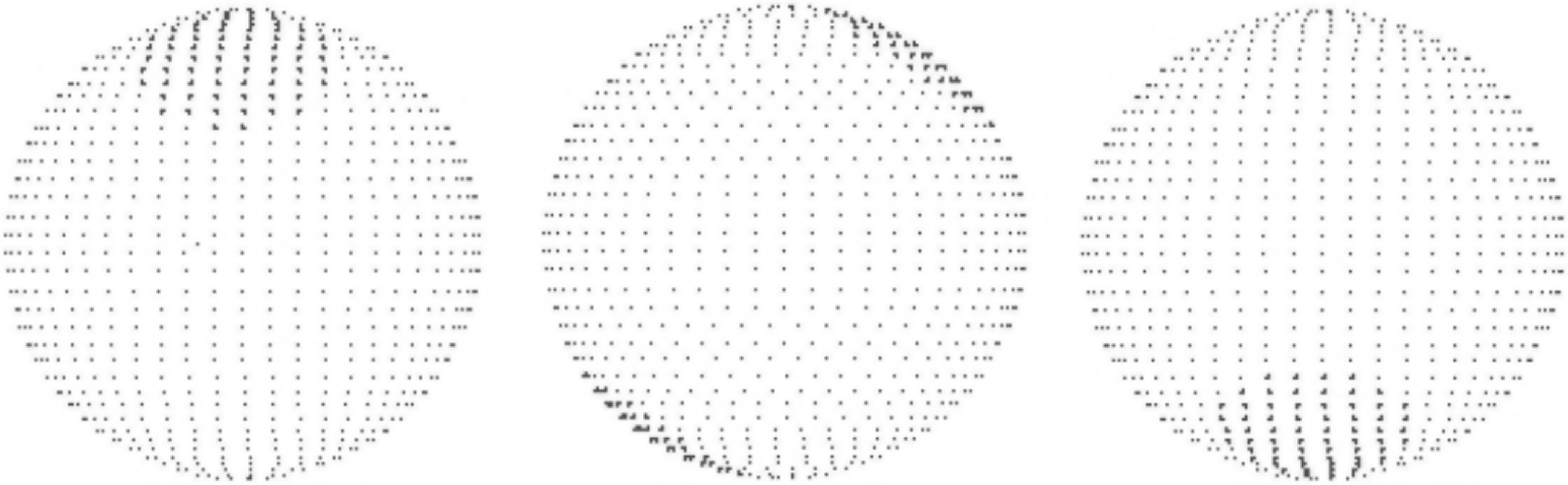}}
\caption{PHOEBE images of spotted star model at phases 0.0, 0.25 and
  0.5, from left to right.}
\label{spotplot}
\end{figure}

\begin{figure}
\resizebox{\hsize}{!}{\includegraphics{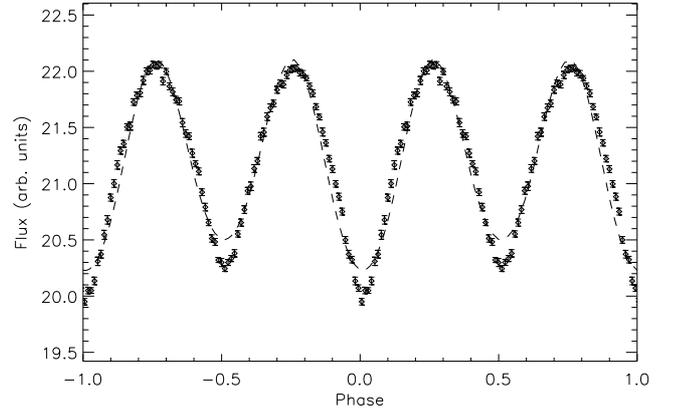}}
\caption{Best light curve fit for PHOEBE spotted star model.}
\label{spotlcfit}
\end{figure}

\begin{figure}
\resizebox{\hsize}{!}{\includegraphics{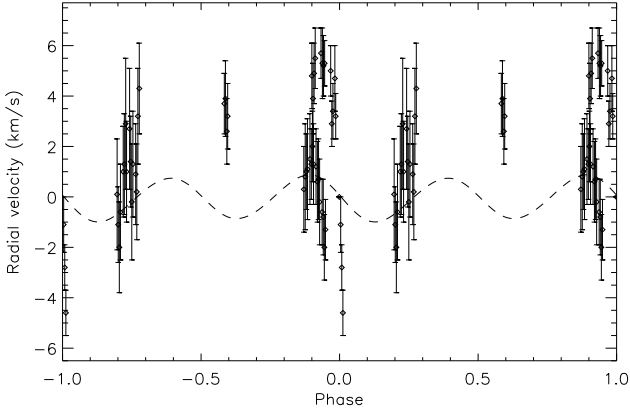}}
\caption{Best RV curve fit for PHOEBE spotted star model.}
\label{spotrvfit}
\end{figure}

Our first model regards J2344 as what its spectrum indicates: a single
mid-K dwarf, rotating with a period of 18\,461.6~s.  The low amplitude
of light curve variability would be consistent with rotational variation
caused by cool surface spots.  However, the alternating deeper and
shallower minima at phases 0.0 and 0.5 (Fig.~\ref{lc}), observed over
many years of ASAS and SuperWASP data, would require two large stable
spots of different areas and/or temperatures, located on
diametrically-opposite sides of the star (Fig.~\ref{spotplot}).  This
could be achieved if the spots were somehow pinned to the star's
magnetic poles (\citet{harrison} claimed similar cool stable polar
spots on many K-class rotational variables observed with Kepler), and
if the star were an oblique rotator \citep{stibbs}, having its
magnetic axis at an angle to its axis of rotation.  The small RV
excursions from zero would then be caused by a form of the
Rossiter-McLaughlin effect \citep{rossiter,mclaughlin} associated with
the spots, as observed by \citet{huber}.  Precession due to the
different alignments of magnetic and rotational axes might explain the
4.19~y meta-cycle of period changes \citep{monaghan}.

Using PHOEBE again to test this idea, we modelled a single rotational
variable in similar fashion to \citeauthor{harrison}, setting the
input orbital period to the assumed rotational period and turning off
the light from the detached companion.  Since we were interested in
reproducing the RVs as well as the light curve, we set the mass ratio
as low as possible so that the modelled curves were both flat before
the introduction of spots.  The (primary) star was given a mass and
effective temperature consistent with a K5V spectrum, and $i$ was set
to $90\degr$ for simplicity.  Two spots were then added to the primary
in accordance with the model, and adjusted manually until reasonable
light and RV curve fits were obtained.  The final spot location
parameters were colatitudes $35\degr$ and $145\degr$ and longitudes
$0\degr$ and $180\degr$ respectively, to simulate stable location on
the poles of a magnetic axis at $35\degr$ to the rotational axis.  One
was given a slightly larger radius ($26\degr$ vs. $25\degr$) to
reproduce the different depths of light curve minima, but both were
set to the same temperature (20\% of average).

Figs.~\ref{spotlcfit} and \ref{spotrvfit} show the resulting fits.
The light curve is fairly well reproduced, both in amplitude and
different minima depths.  The fitted RV curve is somewhat smaller in
amplitude than that observed, but does match some of the velocity
trends in the observed curve, notably the maximum around phase 0.9 and
the rising trend around phase 0.25.  Better fits might be obtained by
modelling non-spherical spots, but these initial results may at least
serve as proof of concept.  A greater problem for the one-star model
is the lack of evidence (to our knowledge) for other low-mass oblique
rotators.

\subsection{Two-star model}

\begin{figure}
\resizebox{\hsize}{!}{\includegraphics{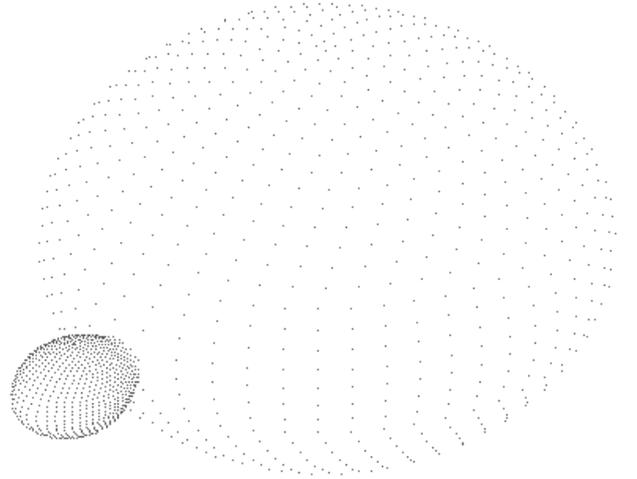}}
\caption{PHOEBE image of K dwarf+brown dwarf binary model at phase 0.9.}
\label{bdplot}
\end{figure}

\begin{figure}
\resizebox{\hsize}{!}{\includegraphics{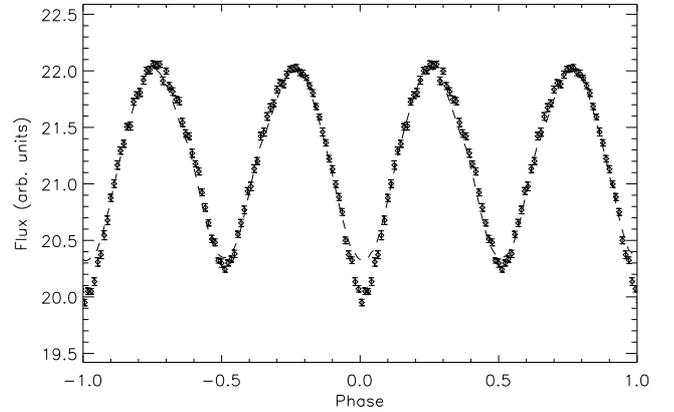}}
\caption{Best light curve fit for PHOEBE K dwarf+brown dwarf binary model.}
\label{bdlcfit}
\end{figure}

\begin{figure}
\resizebox{\hsize}{!}{\includegraphics{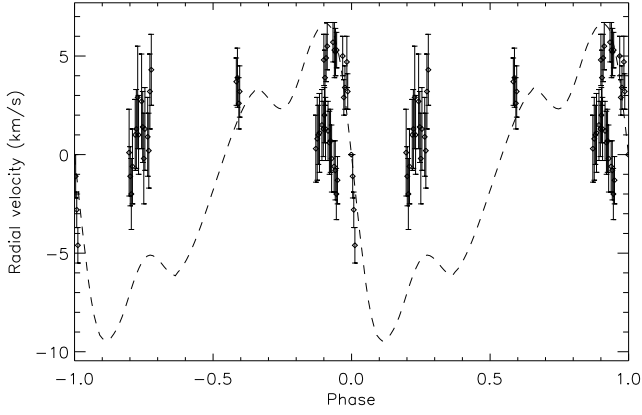}}
\caption{Best primary RV curve fit for PHOEBE K dwarf+brown dwarf
binary model.}
\label{bdrvfit}
\end{figure}

Our second model for J2344 takes the light curve at face value, seeing
it as an eclipsing binary in contact configuration, with a mid-K dwarf
as primary, and a secondary component making a very limited
contribution to the spectrum.  In order to reproduce the observed
amplitude of RV variation (associated with the primary, in this
model), the secondary's mass must be in the brown dwarf range
(Fig.~\ref{bdplot}).  The observed shape of the RV curve would then be
due to the Rossiter-McLaughlin effect as the secondary obscures each
side of the primary in turn, and the 4.19~y meta-cycle of period
changes could be explained by the Applegate mechanism
\citep{applegate}.

A PHOEBE model with $i=59\degr$, $q=0.025$, $M_1=0.79M_{\sun}$ and
$M_2=0.02M_{\sun}$ produced an excellent fit to the observed light
curve amplitude and shape, though the different depths of minima could
not be easily reproduced (Fig.~\ref{bdlcfit}).  The RV fit
(Fig.~\ref{bdrvfit}) was of slightly greater amplitude than the
observed curve, but its Rossiter-McLaughlin effect-induced variations
matched the velocity trends reasonably well, as with the spotted star
model.  We note, however, that PHOEBE is not intended to model planetary-mass
companions, and may not model well objects in the brown dwarf range
either, so these model outputs should be regarded with caution.

\subsection{Three-star model}

\begin{figure}
\resizebox{\hsize}{!}{\includegraphics{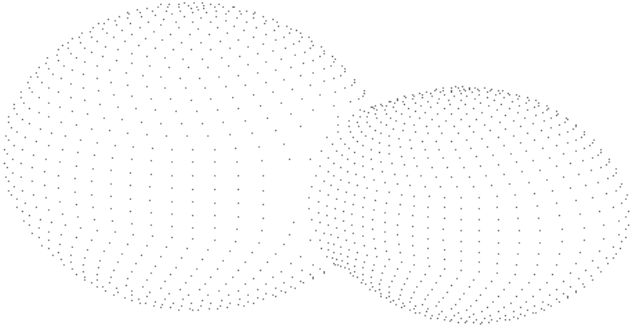}}
\caption{PHOEBE image of M+M dwarf binary model at phase 0.15.}
\label{tripleplot}
\end{figure}

\begin{figure}
\resizebox{\hsize}{!}{\includegraphics{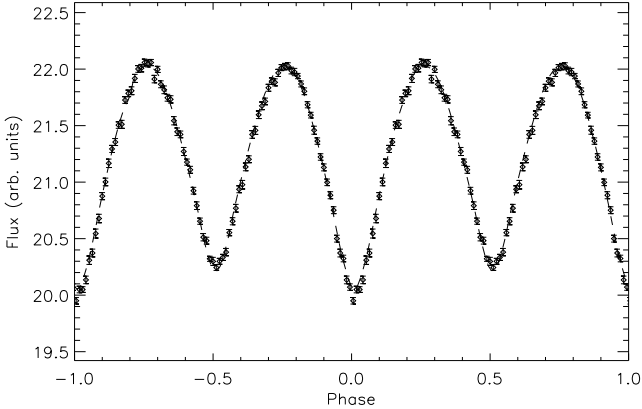}}
\caption{Best light curve fit for PHOEBE M+M dwarf binary model.}
\label{triplelcfit}
\end{figure}

\begin{figure}
\resizebox{\hsize}{!}{\includegraphics{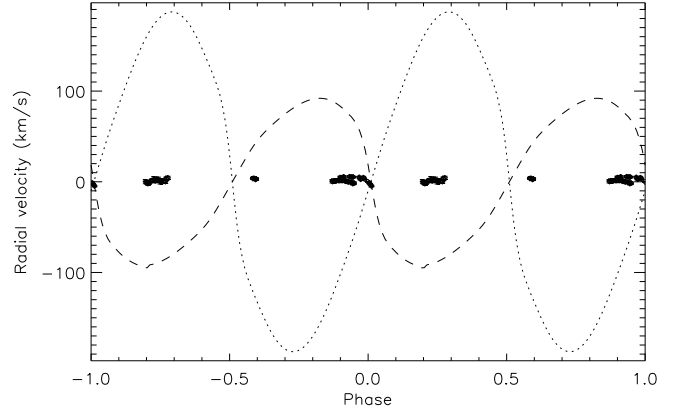}}
\caption{RV curves corresponding to best light curve fit for PHOEBE M+M dwarf binary model (primary curve shown with dashed line; secondary with dotted line.)}
\label{triplervfit}
\end{figure}

Our final model for J2344 is of a triple system, consisting of a very
low-mass contact eclipsing binary orbiting a more massive mid-K star
which dominates the spectrum and obscures the contribution from the
binary.  The light curve is then the sum of a constant flux
contribution from the K star (providing up to 9/11 of the maximum
system flux), and a periodically-variable contribution from the
binary.  The radial velocity curve is almost constant, since it
largely represents the unvarying (on this timescale) position of the K
dwarf.  The 4.19~y cycle of period variation would be a light-time
effect (LITE) resulting from the orbit of the contact binary around
the most massive component of the triple.

A third light can be readily included in PHOEBE's models;
Figs.~\ref{tripleplot} to \ref{triplervfit} show the results of
modelling the eclipsing binary in such a triple system, using
parameters $i=77\degr$, $q=0.5$, $M_1=0.34M_{\sun}$ and
$M_2=0.17M_{\sun}$ (i.e. M dwarfs), and a maximal third light of 18.0
SuperWASP flux units (11.86~$V$ mag).  Fig.~\ref{tripleplot} indicates
the very deep contact required; assuming this, however, an excellent
fit to the light curve is achieved, without even needing spots to be
included for fine-tuning (Fig.~\ref{triplelcfit}).
Fig.~\ref{triplervfit} shows the primary and secondary RV curves
implied for such an eclipsing binary, for reference only, since our
observed SALT velocities are expected to be dominated by the constant
K star, which is not included in the PHOEBE model as a mass, only as a
light source.

Using these masses, totalling $\sim0.5M_{\sun}$, for the binary, and a
plausible $0.65M_{\sun}$ for the K5 system primary, a
binary-to-primary flux ratio of about 1:5 is implied, consistent with
the modelled ratio of around 1:6 given by the SuperWASP light curve
with maximum third light.  We can also insert these values into the
approximate formula for expected LITE changes for a binary in an
edge-on circular orbit with a third body, given by \citet{pribulla} in
their Eq.~5:
\begin{equation}\Delta T \approx \frac{2 M_3
    G^{1/3}}{c}\left[\frac{P_3}{2\pi
    (M_1+M_2)}\right]^{2/3},\end{equation}
where $M_{1,2}$ are the masses of the binary components, $M_3$ is the
mass of the K5 system primary, $P_3$ is the 4.19~y period of $O-C$
oscillations, and $\Delta T$ is the peak-to-peak amplitude of the
$O-C$ variations.  The result is 2640~s, about twice the observed
amplitude (Fig.~\ref{oc}), which is entirely plausible if we do not
expect the angle of inclination to be $90\degr$.  (Indeed, using
\citeauthor{pribulla}'s Eq.~10 with these masses and our observed
semi-amplitude of LITE of 631~s, $i=56\degr$ is suggested.)

\begin{figure}
\resizebox{\hsize}{!}{\includegraphics{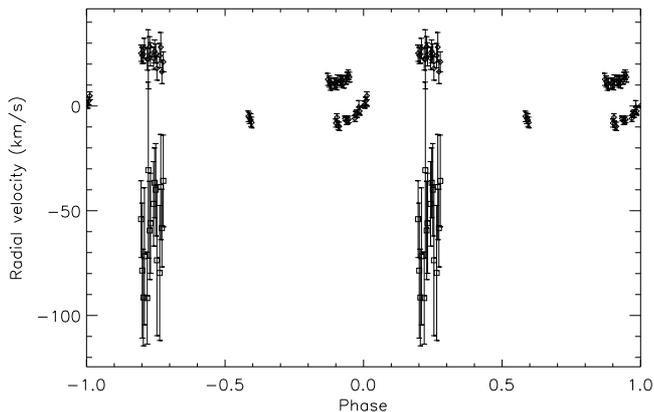}}
\caption{Residual RV curves for J2344, following subtraction of scaled
  K5V template spectrum, and cross-correlation with phase 0 residual
  spectrum.  The stronger cross-correlation peak velocities are shown
  with diamonds; the fainter cross-correlation peak visible near
  quadrature is plotted with squares.}
\label{triplesubrv}
\end{figure}

On the assumption that this model was approximately correct, a
suitably-scaled constant K5 template spectrum was subtracted from each
of our spectra to see whether some trace of an M+M eclipsing binary
spectrum might be detectable in the residuals.  Fig.~\ref{triplesubrv}
shows the resulting RV curves, after cross-correlation with a phase 0
residual spectrum.  A second cross-correlation peak was now marginally
detectable in the spectra near quadrature, yielding RVs in a similar
range, and following a similar upward trend, to those predicted for
the primary curve in Fig.~\ref{triplervfit} near phase 0.25.  The
other RVs reached greater amplitudes than before (Fig.~\ref{rv}) and
might conceivably represent a blending of lines from multiple stellar
components.  Given that our best model for the low-mass eclipsing
binary in this putative triple system involves very deep contact,
it is likely that their lines would be significantly broadened and
blended even if no third spectrum were present to complicate the
picture, making extraction of RVs challenging in any case.

2MASS and WISE colours for the source were also checked for evidence
of an infrared excess which might support the presence of M dwarfs in
the system.  The results were inconclusive, however: 2MASS J$-$H and
H$-$K colours were in ranges expected for a K5 star, while the WISE
colours were inconsistent, possibly being contaminated by nearby
sources.

Although we lack conclusive evidence for it, a triple system seems the
most likely of our three models for J2344.  We have not assessed the
dynamical stability of such a triple, but note the recent detection of
a young hierarchical triple composed of a late-K primary and a pair of
mid-M dwarfs in wide orbits \citep{deacon} which shows some similarity
to the system posited here.  Using the observed and theoretical
absolute magnitudes for the stars in this model, a distance of
80--90~pc is indicated.  The calculated separation between system
primary and contact binary is 2.7~AU, which would then correspond to
an angular separation of $\sim0\farcs03$, making the components
resolvable in principle.  The expected RV amplitude for the system
primary, over a 4.19~y orbital period with the binary, would be around
8~km~s\textsuperscript{-1} (assuming a circular orbit and the same
angle of inclination as for the contact binary), which might also be
detectable in the long term.

\section{Conclusions}

Object J2344, which we originally thought might be an eclipsing binary
close to stellar merger, has proved to be unusual and intriguing in a
different way.  Of the three models investigated here, a triple system
containing a low-mass eclipsing binary appears the most plausible
explanation for the apparently conflicting photometric and
spectroscopic data, and provides an appealing reason for the dramatic
cyclic variations in period length observed as well.  If confirmed, it
should provide a useful contribution to studies of multiple systems,
alongside the quadruple doubly-eclipsing system already detected in
our collection of short-period SuperWASP eclipsing binary candidates
\citep{lohr13}.

Alternatively, if one of our other explanations proves more likely,
J2344 could add to our knowledge of brown dwarfs, or constitute a rare
type of rotational variable.  Still other explanations are no doubt
conceivable e.g. involving higher multiplicity of the system; in any
case, this appears to be an interesting object worthy of further
observation.  We would hope in the future to obtain multi-colour
photometry and near-infrared spectroscopy of J2344, with improved
phase coverage, in the expectation that the greater contrast available
at longer wavelengths would increase the opportunity of detecting
cool, low-mass objects within the system.

\begin{acknowledgements}
The WASP project is funded and operated by Queen's University Belfast,
the Universities of Keele, St. Andrews and Leicester, the Open
University, the Isaac Newton Group, the Instituto de Astrofisica de
Canarias, the South African Astronomical Observatory and by STFC.
Some of the observations reported in this paper were obtained with the
Southern African Large Telescope (SALT) under program 2012-1-UKSC-007
(PI: Andrew Norton).  This work was supported by the Science and
Technology Funding Council and the Open University.
\end{acknowledgements}

\bibliographystyle{aa}
\bibliography{reflist}

\begin{longtable}{r r r}
\caption{\label{mintab} Times of primary eclipses for J2344.}\\
\hline\hline
Epoch & Heliocentric & Uncertainty\\
 & Julian Date & (days)\\
\hline
\endfirsthead
\caption{continued.}\\
\hline\hline
Epoch & Heliocentric & Uncertainty\\
 & Julian Date & (days)\\
\hline
\endhead
$-$2461 & 2453886.60695 & 0.00072\\
$-$2433 & 2453892.58935 & 0.00069\\
$-$2428 & 2453893.65657 & 0.00107\\
$-$2419 & 2453895.58295 & 0.00267\\
$-$2414 & 2453896.64885 & 0.00093\\
$-$2405 & 2453898.57511 & 0.00139\\
$-$2377 & 2453904.55514 & 0.00267\\
$-$2372 & 2453905.62436 & 0.00085\\
$-$2363 & 2453907.54660 & 0.00077\\
$-$2349 & 2453910.53966 & 0.00075\\
$-$2344 & 2453911.60580 & 0.00083\\
$-$2321 & 2453916.52289 & 0.00123\\
$-$2316 & 2453917.58959 & 0.00069\\
$-$2311 & 2453918.65917 & 0.00101\\
$-$2307 & 2453919.51645 & 0.00267\\
$-$2302 & 2453920.58170 & 0.00107\\
$-$2283 & 2453924.64256 & 0.00080\\
$-$2237 & 2453934.47576 & 0.00267\\
$-$2232 & 2453935.53884 & 0.00072\\
$-$2227 & 2453936.60849 & 0.00056\\
$-$2199 & 2453942.59180 & 0.00267\\
$-$2195 & 2453943.44516 & 0.00267\\
$-$2190 & 2453944.51297 & 0.00267\\
$-$2148 & 2453953.48793 & 0.00040\\
$-$2143 & 2453954.55777 & 0.00077\\
$-$2115 & 2453960.53955 & 0.00067\\
$-$2111 & 2453961.39629 & 0.00093\\
$-$2097 & 2453964.38780 & 0.00053\\
$-$2092 & 2453965.45552 & 0.00037\\
$-$2050 & 2453974.42931 & 0.00040\\
$-$2045 & 2453975.49880 & 0.00051\\
$-$2031 & 2453978.48999 & 0.00045\\
$-$2027 & 2453979.34570 & 0.00077\\
$-$2008 & 2453983.40481 & 0.00104\\
$-$1985 & 2453988.31864 & 0.00267\\
$-$1980 & 2453989.38704 & 0.00072\\
$-$1961 & 2453993.44873 & 0.00134\\
$-$1947 & 2453996.44038 & 0.00267\\
$-$1943 & 2453997.29469 & 0.00267\\
$-$1933 & 2453999.43164 & 0.00053\\
$-$1929 & 2454000.28576 & 0.00067\\
$-$1924 & 2454001.35314 & 0.00099\\
$-$1919 & 2454002.42266 & 0.00254\\
$-$1915 & 2454003.27746 & 0.00053\\
$-$1901 & 2454006.26940 & 0.00075\\
$-$1896 & 2454007.33783 & 0.00075\\
$-$1891 & 2454008.40794 & 0.00142\\
$-$1825 & 2454022.50757 & 0.00080\\
$-$1797 & 2454028.49182 & 0.00059\\
$-$1788 & 2454030.41738 & 0.00267\\
$-$1784 & 2454031.27006 & 0.00267\\
$-$1779 & 2454032.33882 & 0.00061\\
$-$1765 & 2454035.33068 & 0.00056\\
$-$1760 & 2454036.39877 & 0.00115\\
$-$1756 & 2454037.25197 & 0.00085\\
$-$1751 & 2454038.32308 & 0.00080\\
$-$1718 & 2454045.37236 & 0.00118\\
$-$1713 & 2454046.44185 & 0.00069\\
$-$1709 & 2454047.29360 & 0.00267\\
$-$1695 & 2454050.28889 & 0.00147\\
$-$1676 & 2454054.34772 & 0.00067\\
$-$673 & 2454268.66929 & 0.00267\\
$-$664 & 2454270.59202 & 0.00093\\
$-$659 & 2454271.66180 & 0.00123\\
$-$650 & 2454273.58502 & 0.00048\\
$-$645 & 2454274.65293 & 0.00037\\
$-$598 & 2454284.70093 & 0.00032\\
$-$584 & 2454287.68879 & 0.00267\\
$-$580 & 2454288.54254 & 0.00267\\
$-$561 & 2454292.60289 & 0.00075\\
$-$552 & 2454294.52532 & 0.00043\\
$-$547 & 2454295.59421 & 0.00053\\
$-$542 & 2454296.66176 & 0.00080\\
$-$538 & 2454297.51815 & 0.00267\\
$-$533 & 2454298.58606 & 0.00072\\
$-$528 & 2454299.65271 & 0.00091\\
$-$524 & 2454300.50250 & 0.00267\\
$-$519 & 2454301.57873 & 0.00134\\
$-$505 & 2454304.56864 & 0.00083\\
$-$384 & 2454330.42290 & 0.00101\\
$-$379 & 2454331.49316 & 0.00099\\
$-$374 & 2454332.55956 & 0.00045\\
$-$370 & 2454333.41443 & 0.00061\\
$-$365 & 2454334.48354 & 0.00048\\
$-$346 & 2454338.54382 & 0.00064\\
$-$318 & 2454344.52434 & 0.00267\\
$-$309 & 2454346.44888 & 0.00061\\
$-$300 & 2454348.37196 & 0.00075\\
$-$290 & 2454350.50762 & 0.00069\\
$-$286 & 2454351.36289 & 0.00067\\
$-$281 & 2454352.43173 & 0.00061\\
$-$276 & 2454353.50031 & 0.00045\\
$-$272 & 2454354.35412 & 0.00048\\
$-$258 & 2454357.34597 & 0.00267\\
$-$244 & 2454360.33693 & 0.00045\\
$-$239 & 2454361.40646 & 0.00048\\
$-$234 & 2454362.47410 & 0.00101\\
$-$230 & 2454363.32844 & 0.00085\\
$-$225 & 2454364.39655 & 0.00077\\
$-$220 & 2454365.46480 & 0.00128\\
$-$188 & 2454372.30561 & 0.00267\\
$-$160 & 2454378.28677 & 0.00061\\
$-$155 & 2454379.35438 & 0.00051\\
$-$127 & 2454385.33677 & 0.00208\\
$-$122 & 2454386.40601 & 0.00059\\
$-$118 & 2454387.26227 & 0.00123\\
$-$108 & 2454389.39839 & 0.00048\\
$-$103 & 2454390.46666 & 0.00099\\
$-$99 & 2454391.31929 & 0.00083\\
$-$94 & 2454392.39075 & 0.00152\\
$-$89 & 2454393.45708 & 0.00069\\
$-$66 & 2454398.37012 & 0.00267\\
$-$61 & 2454399.44210 & 0.00267\\
$-$33 & 2454405.42300 & 0.00208\\
$-$29 & 2454406.27824 & 0.00184\\
$-$24 & 2454407.34657 & 0.00267\\
$-$10 & 2454410.33193 & 0.00534\\
$-$5 & 2454411.40659 & 0.00104\\
0 & 2454412.47430 & 0.00093\\
4 & 2454413.33004 & 0.00222\\
9 & 2454414.39831 & 0.00088\\
14 & 2454415.46708 & 0.00267\\
18 & 2454416.31957 & 0.00267\\
23 & 2454417.39073 & 0.00101\\
28 & 2454418.45679 & 0.00267\\
42 & 2454421.45036 & 0.00267\\
46 & 2454422.30272 & 0.00048\\
51 & 2454423.37228 & 0.00120\\
56 & 2454424.44301 & 0.00120\\
60 & 2454425.29264 & 0.00104\\
79 & 2454429.35473 & 0.00134\\
93 & 2454432.34553 & 0.00101\\
98 & 2454433.42021 & 0.00096\\
1105 & 2454648.58473 & 0.00096\\
1110 & 2454649.65321 & 0.00267\\
1161 & 2454660.54641 & 0.00099\\
1166 & 2454661.61721 & 0.00267\\
1180 & 2454664.61213 & 0.00267\\
1189 & 2454666.53495 & 0.00101\\
1194 & 2454667.60021 & 0.00267\\
1217 & 2454672.51604 & 0.00310\\
1222 & 2454673.58436 & 0.00171\\
1227 & 2454674.65122 & 0.00112\\
1236 & 2454676.57263 & 0.00267\\
1241 & 2454677.64389 & 0.00184\\
1245 & 2454678.49766 & 0.00134\\
1269 & 2454683.62471 & 0.00163\\
1273 & 2454684.48023 & 0.00123\\
1278 & 2454685.55084 & 0.00254\\
1283 & 2454686.62008 & 0.00155\\
1287 & 2454687.47046 & 0.00171\\
1292 & 2454688.53802 & 0.00267\\
1297 & 2454689.60880 & 0.00083\\
1301 & 2454690.46396 & 0.00147\\
1306 & 2454691.53129 & 0.00184\\
1353 & 2454701.57659 & 0.00267\\
1371 & 2454705.42150 & 0.00126\\
1409 & 2454713.54132 & 0.00150\\
1413 & 2454714.39318 & 0.00101\\
1418 & 2454715.46511 & 0.00259\\
1423 & 2454716.52846 & 0.00534\\
1446 & 2454721.44473 & 0.00096\\
1451 & 2454722.51393 & 0.00126\\
1488 & 2454730.41819 & 0.00174\\
1493 & 2454731.48987 & 0.00112\\
1497 & 2454732.34311 & 0.00267\\
1507 & 2454734.47645 & 0.00195\\
1511 & 2454735.33269 & 0.00168\\
1521 & 2454737.46938 & 0.00206\\
1525 & 2454738.32459 & 0.00267\\
1535 & 2454740.46443 & 0.00101\\
1549 & 2454743.45446 & 0.00139\\
1553 & 2454744.30939 & 0.00139\\
1558 & 2454745.37454 & 0.00171\\
1563 & 2454746.44635 & 0.00267\\
1577 & 2454749.43736 & 0.00088\\
1614 & 2454757.34004 & 0.00085\\
1619 & 2454758.41305 & 0.00267\\
1623 & 2454759.26587 & 0.00267\\
1637 & 2454762.25622 & 0.00123\\
1642 & 2454763.32609 & 0.00144\\
1647 & 2454764.39387 & 0.00187\\
1652 & 2454765.45763 & 0.00206\\
1656 & 2454766.31757 & 0.00184\\
1661 & 2454767.38690 & 0.00110\\
1666 & 2454768.45102 & 0.00267\\
1675 & 2454770.37485 & 0.00267\\
1680 & 2454771.44470 & 0.00112\\
1684 & 2454772.30295 & 0.00267\\
1694 & 2454774.43463 & 0.00200\\
1698 & 2454775.28986 & 0.00147\\
1750 & 2454786.40225 & 0.00152\\
2799 & 2455010.54183 & 0.00286\\
2809 & 2455012.68445 & 0.00267\\
2813 & 2455013.53607 & 0.00224\\
2818 & 2455014.60170 & 0.00534\\
2851 & 2455021.65333 & 0.00267\\
2855 & 2455022.50561 & 0.00126\\
2888 & 2455029.56044 & 0.00288\\
2893 & 2455030.62470 & 0.00267\\
2897 & 2455031.48286 & 0.00267\\
2902 & 2455032.55379 & 0.00112\\
2907 & 2455033.61997 & 0.00267\\
2911 & 2455034.47662 & 0.00267\\
2925 & 2455037.46840 & 0.00110\\
2930 & 2455038.53667 & 0.00131\\
2935 & 2455039.60552 & 0.00304\\
2940 & 2455040.66175 & 0.00224\\
2963 & 2455045.59261 & 0.00136\\
2967 & 2455046.43898 & 0.00123\\
2972 & 2455047.51171 & 0.00176\\
3019 & 2455057.55330 & 0.00267\\
3023 & 2455058.41182 & 0.00801\\
3028 & 2455059.47648 & 0.00179\\
3033 & 2455060.54497 & 0.00534\\
3038 & 2455061.61446 & 0.00267\\
3051 & 2455064.38493 & 0.00085\\
3056 & 2455065.46223 & 0.00174\\
3061 & 2455066.52331 & 0.00267\\
3065 & 2455067.38210 & 0.00267\\
3070 & 2455068.45596 & 0.00534\\
3084 & 2455071.44529 & 0.00321\\
3094 & 2455073.57723 & 0.00136\\
3098 & 2455074.43122 & 0.00219\\
3103 & 2455075.49689 & 0.01068\\
3149 & 2455085.32602 & 0.00147\\
3173 & 2455090.46059 & 0.00160\\
3187 & 2455093.44727 & 0.00267\\
3205 & 2455097.29448 & 0.00072\\
3271 & 2455111.40251 & 0.00339\\
3280 & 2455113.31981 & 0.00136\\
3285 & 2455114.39646 & 0.00131\\
3304 & 2455118.45370 & 0.00248\\
3308 & 2455119.30761 & 0.00158\\
3318 & 2455121.43799 & 0.00267\\
3322 & 2455122.29840 & 0.00168\\
3327 & 2455123.36042 & 0.00267\\
3332 & 2455124.42853 & 0.00267\\
3346 & 2455127.42506 & 0.00267\\
3350 & 2455128.27909 & 0.00534\\
3411 & 2455141.31090 & 0.00211\\
3416 & 2455142.38037 & 0.00267\\
3453 & 2455150.29269 & 0.00267\\
3463 & 2455152.42755 & 0.00155\\
3467 & 2455153.28229 & 0.00534\\
6183 & 2455733.65438 & 0.00534\\
6216 & 2455740.69178 & 0.00101\\
6262 & 2455750.50511 & 0.00534\\
6267 & 2455751.59120 & 0.00534\\
6272 & 2455752.65743 & 0.00267\\
6276 & 2455753.49949 & 0.01068\\
6286 & 2455755.66517 & 0.00267\\
6290 & 2455756.48744 & 0.00801\\
6295 & 2455757.56952 & 0.00278\\
6300 & 2455758.64359 & 0.01068\\
6304 & 2455759.49594 & 0.00801\\
6309 & 2455760.56053 & 0.00150\\
6314 & 2455761.63443 & 0.00267\\
6318 & 2455762.48263 & 0.00534\\
6323 & 2455763.55964 & 0.00267\\
6328 & 2455764.63517 & 0.00801\\
6332 & 2455765.47635 & 0.00166\\
6365 & 2455772.53046 & 0.00128\\
6370 & 2455773.59822 & 0.00267\\
6374 & 2455774.44192 & 0.01068\\
6379 & 2455775.53433 & 0.00801\\
8793 & 2456291.33132 & 0.00069\\
\end{longtable}

\end{document}